\documentclass{aa}

\usepackage{graphicx}
\usepackage{xcolor}
\usepackage{txfonts}
\usepackage[utf8]{inputenc}

\begin{document}

\title{Galaxy cluster mass density profile derived using\\the submillimetre galaxies magnification bias}
\titlerunning{Galaxy cluster mass density profile with Magnification Bias}
\authorrunning{Fernandez L. et al.}

\author{Fernandez L.\inst{1}, Cueli M. M.\inst{2,3}, Gonz{\'a}lez-Nuevo J.\inst{2,3}, Bonavera L.\inst{2,3}, Crespo D.\inst{2,3}, Casas J. M.\inst{2,3}, Lapi A.\inst{4,5,6,7}}

\institute{$^1$Mathematisches Institute, Universit\"at Bern, Sidlerstrasse 5, 3012 Bern, Switzerland \\
$^2$Departamento de Fisica, Universidad de Oviedo, C. Federico Garcia Lorca 18, 33007 Oviedo, Spain\\
$^3$Instituto Universitario de Ciencias y Tecnologías Espaciales de Asturias (ICTEA), C. Independencia 13, 33004 Oviedo, Spain\\
$^4$SISSA, Via Bonomea 265, 34136 Trieste, Italy\\
$^5$IFPU - Institute for fundamental physics of the Universe, Via Beirut 2, 34014 Trieste, Italy\\
$^6$IRA-INAF, Via Gobetti 101, 40129 Bologna, Italy\\
$^7$INFN-Sezione di Trieste, via Valerio 2, 34127 Trieste,  Italy\\
}

\date{Received xxx, xxxx; accepted xxx, xxxx}
\abstract
{The magnification bias is a gravitational lensing effect that produces an increase or decrease in the detection probability of background sources near the position of a lense. The special properties of the submillimetre galaxies (SMGs; steep source number counts, high redshift, and a very low cross-contamination with respect to the optical band) makes them the optimal background sample for magnification bias studies.
}
{We want to study the average mass density profile of tens to hundreds of clusters of galaxies acting as lenses that produce a magnification bias on the SMGs, and to estimate their associated masses and concentrations for different richness ranges. The cluster richness is defined as $R=L_{200}/L_*$ with $L_{200}$ as the total r-band luminosity within the radius $r_{200}$.}
{The background sample is composed of SMGs observed by \textit{Herschel} with $1.2<z<4.0$ (mean redshift at $\sim 2.3$) while the foreground sample is made up of galaxy clusters extracted from the Sloan Digital Sky Survey III with photometric redshifts of $0.05< z< 0.8$ (mean redshift at $\sim 0.38$).
Measurements are obtained by stacking the SMG--cluster pairs to estimate the cross-correlation function using the Davis-Peebles estimator. This methodology allows us to derive the mass density profile for a wide range of angular scales, $\sim 2-250$ arcsec or $\sim 10-1300$ kpc for $z=0.38$, with a high radial resolution, and in particular to study the inner part of the dark matter halo ($<100$ kpc). In addition, we also divide the cluster sample into five bins of richness and we analyse the estimated cross-correlation data using different combinations of the most common theoretical mass density profiles.
}
{It is impossible to fit the data with a single mass density profile at all scales: in the inner part there is a clear excess in the mass density profile with respect to the outer part that we interpret as the galactic halo of the big central galaxy. As for the outer part, the estimated average masses increase from $M_{200c}=5.8 \times 10^{13}M_\odot$ to $M_{200c}=51.5\times 10^{13} M_\odot$ ($M_{200c}=7.1\times 10^{13} M_\odot$ for the total sample). With respect to the concentration parameter, its average also increases with richness from $C=0.74$ to $C=1.74$ ($C=1.72$ for the total sample). In the small-scale regions, the obtained average masses fluctuate around $M_{200c}=3-4 \times 10^{13}M_\odot$ with average concentration values of around $C\sim4$.}
{The total average masses are in perfect agreement with the mass--richness relationship estimated from the cluster catalogue. In the bins of  lowest richness, the central galactic halo constitutes $\sim 40$\% of the total mass of the cluster and its relevance decreases for higher richness values. While the estimated average concentration values of the central galactic halos are in agreement with traditional mass--concentration relationships, we find low concentrations for the outer part. Moreover, the concentrations decrease for lower richness values, probably indicating that the group of galaxies cannot be considered to be relaxed systems. Finally, we notice a systematic lack of signal at the transition between the dominance of the cluster halo and the central galactic halo ($\sim 100$ kpc). This feature is also present in previous studies using different catalogues and/or methodologies, but is never discussed.
}


\maketitle
   
\section{Introduction}

The magnification bias \citep{Sch92} is a gravitational lensing effect that consists of an increase or decrease in the detection probability of background sources near the positions of  lenses. It is due to a modification of the integrated source number counts of the background objects, which produces an excess or lack of sources at a given flux density limit, and is mainly related to the logarithmic slope of the integrated number counts ($\beta$; $n(>S)= A S^{-\beta}$). Indeed, very steep source number counts ($\beta>2$) enhance the effect of a magnification bias, making the more frequent (though less easily identified) weak lensing events more easily detectable. We can estimate the magnification bias through the non-zero signal that is produced in the cross-correlation function (CCF) between two source samples with non-overlapping redshift distributions \citep{Scr05,Men10,Hil13,Bar01}.

Not only do the submillimetre galaxies (SMGs) discovered within the
\textit{Herschel} Astrophysical Terahertz Large Area Survey
\citep[H-ATLAS;][]{Eal10} data have steep source number counts, $\beta > 3$, but
many of them are also  at high redshift, $z>1$, and show very low
cross-contamination with respect to the optical band (i.e. the foreground lens
is `transparent' at submillimetre wavelengths and the background source is
invisible in the optical band). These properties make SMGs promising candidates
to be successfully used as a background sample for magnification bias studies,
as in \citet{GON14,GON17} where the CCF was measured with high significance, and
as directly observed by \citet{DUN20} with the Atacama Large Millimetre Array
(ALMA). Moreover, the SMGs are used to investigate the halo mass, projected mass
density profile, and concentration of foreground samples of quasi-stellar
objects \citep[QSOs; ][]{BON19}, for cosmological studies \citep{BON20, GON21, BON21} and to observationally constrain the halo mass function \citep[][Cueli et al., in prep.]{CUE21}.

On the other hand, galaxy clusters are massive bound systems usually placed in the knots of filamentary structures and are used to track the large-scale structure of the Universe.
They are being exploited for cosmological studies \citep[e.g. ][]{ALL11}  exploring the evolution of galaxies \citep[][]{DRE80,BUT78,BUT84,GOT03} and the lensed high-redshift galaxies \citep[e.g.][]{BLA99}. Moreover, clusters have been correlated with background objects to investigate the potential lensing effects \citep[][]{MYE05,LOP08}.

Stacking techniques are often used when the signal to be detected is faint but highly probable, as in the case of weak lensing. These methods allow a statistical study of the overall signal by co-adding the emission from many weak or undetected objects, because single weak lensing events are hardly detectable in general.
Some examples of the applications of stacking techniques are: exploiting \textit{Planck} data to recover the very weak  integrated signal of the Sachs–Wolfe effect  \citep{Pla14,Pla16b}, studying the faint polarised signal of radio and infrared sources in the NRAO VLA Sky Survey (NVSS) and \textit{Planck} \citep[see][]{Stil14, BON17a, BON17b}, obtaining the mean spectral energy distribution of optically selected quasars \citep{Bia19}, detecting weak gravitational lensing of the cosmic microwave background in the \textit{Planck} lensing convergence map \citep{Bia18}, and probing star formation in dense environments of $z\sim 1$ lensing haloes \citep{Wel16}.

In addition, \cite{UME16} estimated the average surface mass density profile of an X-ray-selected subsample of galaxy clusters by stacking their individual profiles. They found that the stacked density profile is well described by the Navarro–Frenk–White (NFW), Einasto, and DARKexp models and that cuspy halo models with a large-scale two-halo term improve the agreement with the data. In particular, a concentration of $C_{200c} = 3.79^{+0.30}_{-0.28}$ at M$_{200c} = 14.1^{+1.0}_{-1.0} M_{\odot}$ is found for the NFW halo model.

In this work, we apply the stacking technique to obtain the mass density profile of galaxy clusters. In particular, the paper is organised as follows. Section \ref{sec:data} describes the data, while section \ref{sec:method} gives details of the methodology applied for the stacking and CCF estimation. Section \ref{sec:denprof} addresses the theoretical framework for the CCF, weak gravitational lensing, and halo density profiles. Our results and conclusions are presented in Sects. \ref{sec:results} and \ref{sec:concl}.

A flat $\Lambda$ cold dark matter ($\Lambda$CDM) cosmology has been adopted throughout the paper, with the cosmological parameters estimated by \cite{PLA18_VI} ($\Omega_m$ = 0.31, $\sigma_8$ = 0.81 and $h = H_0 /100$ $km$ $s^{-1} Mpc^{-1} = 0.67$).

\section{Data}
\label{sec:data}
\begin{figure}[ht]
 \includegraphics[width=0.49\textwidth]{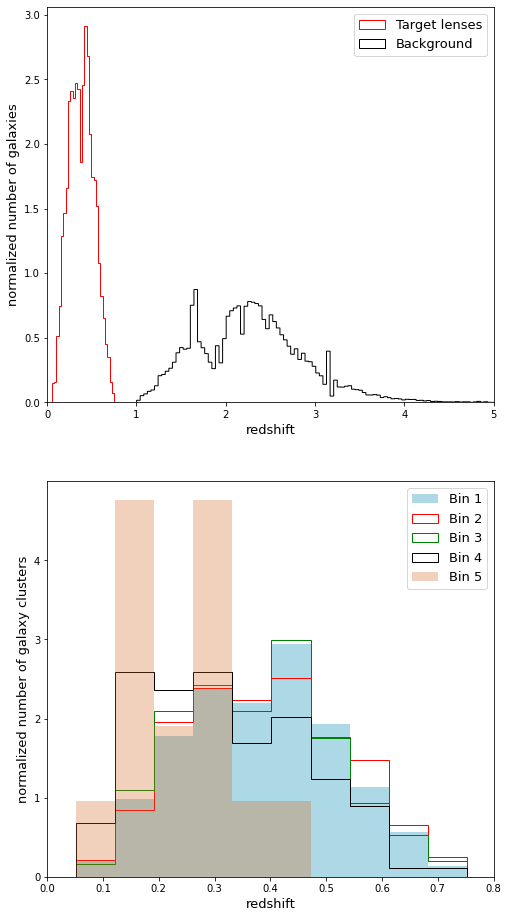} 
 \caption{Top: Redshift distribution of the target lenses from the \citet{WEN12} catalogue (in red) and the background sources from the H-ATLAS sample (in black). Bottom: Redshift distribution of the galaxy clusters for each of the five richness bins.}
 \label{fig:histograms}
 \end{figure}

The background sample consists of the officially detected galaxies in the three H-ATLAS \citep{Pil10} GAMA fields from the first data release (DR1) \citep[][in the equatorial regions at 9, 12, and 14.5 h]{Val16,BOU16,Rig11,Pas11,Iba10} and the field centred at the North Galactic Pole \citep[NGP,][]{Smi17,MAD18} from DR2.
In both H-ATLAS DRs there is an implicit $4\sigma$ detection limit at 250 $\mu$m ($\sim S_{250} > 29$ mJy) \citep{Val16,MAD18} and a $3\sigma$ limit at 350 $\mu$m has been applied to increase the robustness of the photometric redshift estimation \citep[as in][]{GON17}.
In addition, we select sources with a photometric redshift $z>1$ in order to avoid any overlap in the redshift distribution of lenses and background sources (see top panel of Fig. \ref{fig:histograms}). 

The photometric redshifts were estimated by means of a minimum $\chi^2$ fit of a template spectral energy distribution (SED) to the Spectral and Photometric Imaging REceiver \citep[SPIRE;][]{GRI10} data \citep[using Photodetector Array Camera and Spectrometer, PACS,][data when possible]{POG10} . It was shown that a good template is the SED of SMM J2135-0102 (`The Cosmic Eyelash' at $z = 2.3$; \cite{Ivi10,Swi10}), which was found to be the best overall template with $\Delta z/(1 + z) = -0.07$ and a dispersion of 0.153 \citep{Ivi16,GON12,Lap11}.
We are finally left with 70707 sources that constitute approximately  29 $\%$ of the initial number of sources.
The redshift distribution of the background sample is shown in Fig. \ref{fig:histograms} (top panel, black line). The mean redshift of the sample is $\left< z\right> = 2.3_{-0.5}^{+0.4}$ (the uncertainty indicates the $1\sigma$ limits). 
The potential effect of blazars or local galaxy interlopers is considered negligible: the number of detectable blazars is completely negligible while the local galaxies would have photometric redshifts much lower than 1 or, even in the event of a catastrophic photometric redshift failure with resolved individual star-forming regions with abnormal temperatures, they will have redshifts lower than the clusters themselves \citep[see ][for more details]{GON10,Lap11,GON12,LOP13}.

As for the potential lenses, the galaxy cluster sample has been extracted from the catalogue presented in \citet{WEN12} (hereafter WHL12), which contains $132684$ galaxy clusters from the Sloan Digital Sky Survey III (SDSS-III) with given photometric redshifts in the range of $0.05\leq z< 0.8$. 
We select those objects corresponding to the NGP region and the three H-ATLAS GAMA fields. This leads to a total of $3651$ galaxy clusters, which constitute our sample of target lenses. Figure \ref{fig:histograms} (top) shows in red the redshift distribution of the foreground sources. The mean redshift of the sample is $\langle z\rangle=0.38$.

Furthermore, following \cite{BAU14}, we divide the galaxy clusters into five bins according to the richness information provided in WHL12. The cluster richness estimated by WHL12 is defined as $R=L_{200}/L_*$ with $L_{200}$ as the total r-band luminosity within the radius $r_{200}$ (the radius where the mean density of a cluster is 200 times the critical density of the Universe). $L_*$ is the evolved characteristic luminosity of galaxies in the r-band, defined as $L_*(z)=L_*(z=0)10^{0.4Qz}$, adopting a passive evolution with $Q=1.62$ \citep{Bla03}.
Table \ref{tab:richness} shows the number of target lenses in each bin and the associated richness range. The redshift distributions of the richness subsamples are depicted and compared in Fig. \ref{fig:histograms} (bottom panel).

\begin{table}[h]
    \centering
    \caption{Cluster sample information for different richness ranges.}
    \begin{tabular}{ccrr}
    \hline\hline
    Bin number&Richness& \# Targets & \# CG pairs\\
    \hline
    Total& 12-220 & 3651 & 11789\\
    1& 12-17&1977 & 6158\\
    2& 18-25& 1102 & 3723\\
    3&26-40& 430 & 1427\\
    4&41-70& 127 & 424\\
    5&71-220& 15 & 57\\
    \hline\hline
    \end{tabular}
    \label{tab:richness}
    \tablefoot{Richness subdivision of the cluster sample with the number of targets in each richness bin and the number of cluster--galaxy pairs. The richness ranges are chosen following \citet{BAU14}.}
\end{table}

\section{Measurements}
\label{sec:method}

\subsection{Stacking}
\label{sec:stack}
Stacking is a statistical method that proves useful when the desired signal is frequent but too weak: by adding up many regions of the sky centred in previously selected positions \citep[see][]{Dol06,Mar09, Bet12}, the signal is enhanced.
In this way, overall statistical information might be obtained when the single event does not have a sufficiently high signal-to-noise ratio (S/N) to be detected.

Similar to what is done in \cite{BON19}, the signal of interest in this work is the cross-correlation due to magnification bias: the excess of detected sources in the background within a certain angular separation with respect to the random scenario. It should be stressed that, as we are looking for the number of background sources near the lens positions, we are stacking at the position of the sources, not their flux densities. This is a very similar approach to the traditional cross-correlation function (CCF) estimator with the additional advantage that it accounts for positional errors and identifies the foreground--background pairs in the stacked map. 

 \cite{BON19}  derived the stacked magnification bias of lensed SMGs in lens positions signposted by QSOs. In this case, we study the stacked magnification bias produced by clusters acting as lenses on background SMGs. Given a lens, we search for background sources within a circular region centred on its position and within an angular radius of $250$ arcsec. In this way, we obtain a map of 500 $\times$ 500 pixels (the chosen pixel size is 1 arcsec) centred at the target position (the position of the brightest cluster galaxy, or BCG, of each galaxy cluster) containing the nearest background sources of the lens, hereafter referred to as cluster--galaxy (CG) pairs.

This procedure is repeated for all the clusters in the target sample and all the maps are then added to obtain the final stacked map, which is normalised to the number of clusters (the 3651 total targets). Then, a $\sigma=2.4$ arcsec Gaussian filter is applied to the map in order to take into account the positional accuracy present in the catalogues. As the cluster centre positional uncertainty is negligible compared to that of the SMGs (the SDSS positional accuracy is better than 0.1 arcsec), the selected $\sigma$ value corresponds to the positional uncertainty estimated for the H-ATLAS catalogues \citep[][]{BOU16,MAD18}. 
The smoothing step is equivalent to substituting a single pixel at the position of every CG pair for a 2D isotropic Gaussian centred on that pixel to take into account the positional uncertainty (i.e. the fact that the background galaxy could not be exactly at the position that appears in the catalogue). This additional step is taken because random positional displacements toward the lens would produce higher excess probabilities than in the opposite direction, introducing an observational bias (different mass density shape or concentration, and therefore mass) that is more important at the smallest angular scales. 

\begin{figure}[t]
 \includegraphics[width=0.49\textwidth]{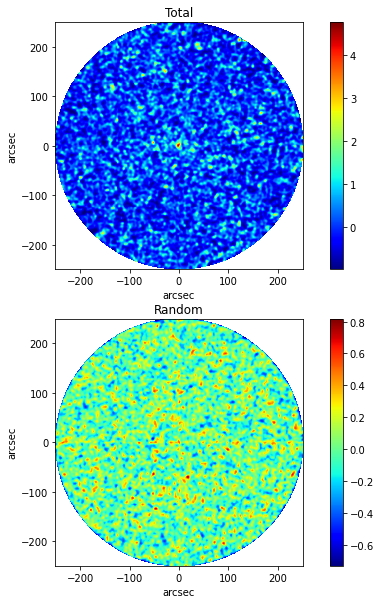}
 \caption{Bottom: Relative excess probability (stacked image/mean - 1) of random pairs placing random targets in the centre and considering the background sources within an angular radius of $250$ arcsec from the position of the target. The pixel size is $1$ arcsec and we apply a $2.4\sigma$ beam Gaussian filter to take into account the positional uncertainties (see text for more details). The mean and standard deviation per pixel of the random stacked image are $1.5\times10^{-5}$ and $2.5\times10^{-6}$.
 Top: Relative excess probability (stacked image/random mean - 1) for the actual CG pairs using the same radius and pixel size as for the random case and smoothed with the same Gaussian filter.}
  \label{fig:totaldensitymap}
\end{figure}

The resulting map with the identified CG pairs is plotted in the top panel of Fig. \ref{fig:totaldensitymap}. The corresponding bottom panel shows the expected signal in the absence of lensing: random cluster positions are simulated and the corresponding distribution maps of CG pairs are produced. To these maps, we apply the same procedure as that applied to the data. As in \citet{BON19}, we simulate ten times the total number of targets, namely 36510, to obtain a homogeneous random map (as we average the results over the number of targets, their total number becomes irrelevant, apart from when calculating statistical uncertainties).
For both panels of Fig. \ref{fig:totaldensitymap}, we decided to use a colour scale that represents the relative excess probability with respect to the random mean value (stacked pairs/ random mean -1) of finding a CG pair. The mean and standard deviation per pixel of the random stacked image are $1.5\times10^{-5}$ and $2.5\times10^{-6}$.

By comparing both images we can extract some preliminary conclusions. 
There is an excess of CG pairs with respect to the random alignment case (at least twice as probable or more than five times the expected random statistical deviation, 0.16), especially in the region located at the centre where a much higher probability is clearly shown (a peak value of relative excess of 4.76 or about 30 times the random statistical deviation), corresponding to a larger lensing effect. As discussed in more detail below, even if most of our signal is in general produced by weak lensing, this stronger excess of CG pairs below 10-20 arcsec is due to the strong lensing effect. At larger angular distances, the distribution of CG pairs is almost isotropic, even if not completely homogeneous. Moreover, its intensity tends to decrease towards the border as expected.

\begin{figure}[t]
 \includegraphics[width=0.49\textwidth]{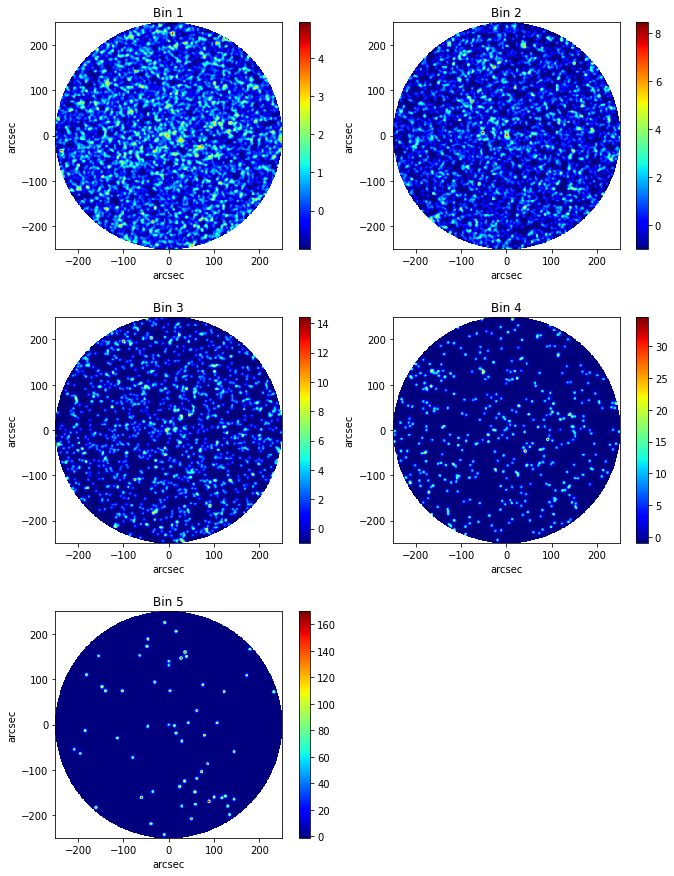}
 \caption{Relative excess probability (stacked image/random mean - 1) of CG pairs for each richness bin, in analogy to the total case in Fig. \ref{fig:totaldensitymap}. The maps are for bins 1 to 5, in the panels from left to right, top to bottom. The colour bars have the same meaning as in Fig. \ref{fig:totaldensitymap}.}
 \label{fig:bindensitymap}
\end{figure}

Furthermore, we repeat the same procedure but considering only the galaxy clusters in each of the richness bins that were described earlier. The relative excess probability images of stacked CG pairs are shown in Fig. \ref{fig:bindensitymap}. 
The number of CG pairs is typically more than three times the number of targets (see Table \ref{tab:richness}). As expected, this number decreases proportionally to the number of targets in each richness bin. We find a marginal tendency of more massive clusters to have more pairs. 

For the bins with the lowest richness, and therefore a larger number of targets and CG pairs, the images are more or less similar to the total case. 
In fact, bins 1, 2, and 3 show a higher density in the centre and an almost isotropic distribution of CG pairs at an angular distance $<250$ arcsec. In bins 4 and 5, the distribution is very discrete due to the poorer statistics (see the number of targets and CG pairs in Table \ref{tab:richness}). As already discussed for the total case, most of the signal is produced by the weak lensing effect, except at angular scales lower than 10-20 arcsec, where the strong lensing effect causes the higher density region located close to the centre of all the bins. In bin 4, and especially bin 5, this effect is less evident, which is expected given the lower statistics and the fact that strong lensing is a rare event. However, it is remarkable that even for bin 5 ---the one with the highest richness and just 15 galaxy clusters---, we have a reasonable number of pairs. Indeed, with $57$ CG pairs, this means that each target galaxy cluster has more than 3 CG pairs on average.

\subsection{Estimation of the cross-correlation function}
\label{sec:xcorr}

We analyse the stacked images through the measurement of the CCF, which allows us to extract physical information about the average properties of the lensing system in general and the lens sample in particular. For this reason, we estimate the CCF using the stacked CG pair maps as in \cite{BON19} instead of applying the traditional methodology as in \cite{GON14,GON17}. 

We draw a finite set of concentric circles centred at the central position of the images. The radii increase logarithmically in steps of 0.05, starting with $1$ arcsec (the first measurements are limited by the pixel size). This defines one initial circle and a set of rings. The pixel values in each circular annulus are added up as DD. The same procedure is applied to the random map (RR). The standard estimator \citep{Dav83} is then calculated as:
\begin{equation}
    \tilde{w}_{x}(\theta)=\frac{\text{DD}}{\text{RR}}-1
    \label{eq:wx}
.\end{equation}
The errors on $\tilde{w}_{x}$ were calculated in two steps: first, we divided each ring into $15$ equal sections, except for the first four rings where there are no more than eight pixels (in these cases, we divided the rings into fewer sections). We then applied a Jackknife method as in \citet{BON19} to estimate the uncertainties for $\text{DD}$ and $\text{RR}$. Finally, we used error propagation according to eq. \eqref{eq:wx}.

The estimated CCFs are shown in Figs. \ref{fig:SISSA_plots} and \ref{fig:NFW_plots} (black points) for the five richness bins (numbered 1 to 5, from left to right, top to bottom) and the total case (bottom right panel).
With the current stacking configuration, we obtain CCFs from 2 to 250 arcsec. These angular scales correspond to $\sim 10$ kpc to $\sim 1.3$ Mpc for $z=0.38$. It should be noted that, by using a single method, we are able to study the mass density profile in a wide spectrum of physical scales. This is a interesting novel characteristic of this methodology with respect to the strong and weak lensing analysis of individual clusters that have to be combined to cover a comparable angular scale range: strong lensing can only be measured in the central part of the clusters where the strong lensing features are produced and can be measured, while the weak lensing cannot be measured towards the most central part of the cluster because of the influence of central galaxies.

In addition, the CCFs confirm the conclusions derived preliminarily from the images: a stronger signal at the smallest angular separations that decreases logarithmically towards the largest. In addition, the maximum of the CCFs increases with richness (mass) as expected for an event related to gravitational lensing. As is clear from the measurements, the CCF of the bins with the  highest richness (4 and 5) shows strong oscillations due to the lack of CG pairs. We discuss this matter in more detail in section \ref{sec:feature}.

%
%--------------------------------
%--------------------------------
%Section 4. Theoretical framework 
%--------------------------------
%--------------------------------
\section{Theoretical framework}
\label{sec:denprof}

In order to extract important physical information about the mass distribution in a galaxy cluster halo from the measured magnification bias, we need to rely on a theoretical framework that firstly connects the observed cross-correlation function with the gravitational lensing amplification, and secondly relates such amplification with a mass density profile. The analysis will then consist of the determination of the mass density profile (or a combination of two profiles) that better explains the observations and can be used to decipher  the physical  halo  characteristics of such a profile, such as mass or concentration.
%-----------------------------------------
%4.1. Gravitational lensing and the cross-correlation function
%--------------------------------------------------------------
\subsection{Gravitational lensing and the cross-correlation function}
\label{sec:theory}

Let 
\begin{equation}
    n_0(>S,z)\equiv \int_S^{\infty} \frac{dN}{dSdz\,d\Omega}dS
\end{equation}
denote the unlensed integrated background source number counts, $n_0$, which is the number of background sources per solid angle and redshift with observed flux density larger than $S$ in the absence of gravitational lensing. Due to the influence of foreground lenses, this quantity is modified at every angular position in the image on account of two separate effects, namely a magnification that allows fainter sources to be observed and a dilution that enlarges the solid angle subtended by the sources in question. More precisely, at an angular position $\vec{\theta}$ within an image, we have \citep{Bar01}
\begin{equation}
    n(>S,z;\vec{\theta})=\frac{1}{\mu(\vec{\theta})}n_0\Big(>\frac{S}{\mu(\vec{\theta})},z \Big),\label{numbercounts1}
\end{equation}
where $\mu(\vec{\theta})$ is the magnification field at angular position $\vec{\theta}$.

Assuming a redshift-independent power-law behaviour of the unlensed integrated number counts, that is, $n_0(>S,z)=AS^{-\beta}$,
\eqref{numbercounts1} becomes 
\begin{equation}
    \frac{n(>S,z;\, \vec{\theta})}{n_0(>S,z)}=\mu^{\,\beta-1}(\vec{\theta}).
\end{equation}

As we aim to relate the magnification field to a direct observable based on galaxy counting, one can interpret that, from the point of view of a lens at a certain redshift $z_d$, the quantity $n(>\!\!S,z_s;\vec{\theta})/n_0(>\!\!S,z_s;\vec{\theta})$ represents the excess (or lack) of background sources (in direction $\vec{\theta}$ relative to the lens) at redshift $z_s>z_d$ with respect to what would be expected in the absence of lensing. Indeed, the angular cross-correlation function between a set of foreground lenses at redshift $z_d$ and a set of background sources at redshift $z_s$ is defined as
\begin{equation}
w_x(\vec{\theta};z_d,z_s)\equiv \langle\delta n_f(\vec{\phi},z_d)\,\delta n_b(\vec{\phi}+\vec{\theta},z_s)\rangle,
\end{equation}
where $\delta n_f$ is the foreground galaxy density contrast, which is due to pure clustering, and 
\begin{equation}
    \delta n_b(\vec{\theta},z)=\frac{n(>S,z;\vec{\theta})}{n_0(>S,z;\vec{\theta})}-1=\mu^{\,\beta-1}(\vec{\theta})-1
,\end{equation}
is the background galaxy density contrast, which is due to magnification. As we are stacking the lenses at a fixed position, which we take as the origin, we have
\begin{equation}
    w_x(\vec{\theta},z_d,z_s)=\mu^{\,\beta-1}(\vec{\theta})-1.\label{crossmag}
\end{equation}

The magnification field can in turn be written in terms of the convergence ($\kappa$) and the shear $(\vec{\gamma})$ fields, which describe the local matter density and the tidal gravitational field, respectively. Indeed, we have \citep{Bar01}
\begin{equation}
    \mu(\vec{\theta})=\frac{1}{(1-\kappa(\vec{\theta}))^2-|\vec{\gamma}(\vec{\theta)}|^2}.\label{magkapgam}
\end{equation}

Therefore, plugging \eqref{magkapgam} into \eqref{crossmag} yields a relation between the angular cross-correlation function and the convergence and shear fields, which are determined by the mass density profile of the lens.

%--------------------------
%4.2. Mass density profiles
%--------------------------
\subsection{Mass density profiles}
\label{sec:profiles}

Let us assume that a lens at an angular diameter distance $D_d$ from the observer deflects the light rays from a source at an angular diameter distance $D_s$. If $\vec{\theta}=\vec{\xi}/D_d$ denotes the angular position of a point on the image plane, then the convergence at that point, $\kappa(\vec{\theta})$, is defined as a dimensionless surface mass density, that is,

\begin{equation}
    \kappa(\vec{\theta})\equiv\frac{\Sigma(D_d\vec{\theta})}{\Sigma_{\text{cr}}},
\end{equation}
where $\Sigma(\vec{\xi})$ is the mass density projected onto a plane perpendicular to the incoming light ray, and 
\begin{equation}
    \Sigma_{\text{cr}}=\frac{c^2}{4\pi G}\frac{D_s}{D_dD_{ds}}
\end{equation}
is the so-called critical surface mass density, where $D_{ds}$ is the angular diameter distance from the lens to the background source.

If a lens is axially symmetric, that is, if $\Sigma(\vec{\xi})=\Sigma(\xi)$, then choosing the symmetry centre as the origin, we have $\kappa(\vec{\theta})=\kappa(\theta)$ and the magnification field is given by
\begin{equation}\label{eq:mu}
    \mu(\theta)=\frac{1}{(1-\bar{\kappa}(\theta))(1+\bar{\kappa}(\theta)-2\kappa(\theta))},
\end{equation}
where $\bar{\kappa}(\theta)$ is the mean surface mass density inside the angular radius $\theta$.

\subsubsection{Navarro-Frenk-White profile}
Let us now assume that the mass of the lens is dominated by dark matter. The best known model for describing its mass density is the Navarro-Frenk-White (NFW) profile \citep{NAV96},
\begin{equation}
    \rho_{\text{NFW}}(r;r_s,\rho_s)=\frac{\rho_s}{(r/r_s)(1 + r/r_s)^2},
\end{equation}
where $r_s$ and $\rho_s$ are the scale radius and density parameters, respectively. If we identify halos at redshift $z$ with spherical regions with a mean overdensity of $200\rho_c(z)$, with $\rho_c(z)$ the critical density of the Universe at redshift $z$, then
\begin{equation}
    \frac{\rho_s}{\rho_c(z)}=\frac{200}{3}\frac{C^3}{\ln{(1+C)}-C/1+C},
\end{equation}
where $C=C(M_{200c},z)\equiv R_{200}/r_s$ is the mean concentration of a halo of mass $M_{200c}$ identified at redshift $z$, and $R_{200}$ is its radius. Throughout the paper, in order to avoid confusion between the different kinds of masses, we use $M_{\text{NFW}}$ to indicate the NFW $M_{\text{200c}}$ mass of a halo.

This profile satisfies \citep{SCH06}:
\begin{equation}
    \kappa_{\text{NFW}}(\theta)=\frac{2r_s\rho_s}{\Sigma_{\text{cr}}}f(\theta/\theta_s)\quad\quad\quad \bar{\kappa}_{NFW}(\theta)=\frac{2r_s\rho_s}{\Sigma_{cr}}h(\theta/\theta_s),
\end{equation}
where $\theta_s\equiv r_s/D_d$ is the angular scale radius,
\begin{equation}
    f(x)\equiv \begin{cases} \frac{1}{x^2-1}-\frac{\arccos{(1/x)}}{(x^2-1)^{3/2}}\quad\quad&\text{if } x>1\\
    \,\,\frac{1}{3} \quad\quad&\text{if } x=1\\
    \frac{1}{x^2-1}+\frac{\text{arccosh}(1/x)}{(1-x^2)^{3/2}} \quad\quad&\text{if } x<1
    \end{cases}
\end{equation}
and
\begin{equation}
    h(x)\equiv \begin{cases} \frac{2}{x^2}\Big(\frac{\arccos{(1/x)}}{(x^2-1)^{1/2}}+\log{\frac{x}{2}}\Big)\quad\quad&\text{if } x>1\\
    \,{\scriptstyle 2\,(1-\log{2})} \quad\quad&\text{if } x=1\\
    \frac{2}{x^2}\Big(\frac{\text{arccosh }(1/x)}{(1-x^2)^{1/2}}+\log{\frac{x}{2}}\Big) \quad\quad&\text{if } x<1
    \end{cases}.
\end{equation}

\subsubsection{Singular isothermal sphere profile}
Another option for parametrising the halo density profile is the singular isothermal sphere (SIS) profile, given by
\begin{equation}
    \rho_{\text{SIS}}=\frac{\sigma_v^2}{2\pi Gr^2},
\end{equation}
which corresponds to a system of particles whose velocity distribution at every radius follows a Maxwell-Boltzmann law with one-dimensional velocity dispersion $\sigma_v$. This profile satisfies \citep{SCH06}
\begin{equation}
    \kappa_{\text{SIS}}=\frac{\theta_E}{2|\theta|}\quad\quad\bar{\kappa}_{\text{SIS}}(\theta)=\frac{\theta_E}{|\theta|},
\end{equation}
where
\begin{equation}
    \theta_E=4\pi\,\bigg(\frac{\sigma_v}{c}\bigg)^2\frac{D_{ds}}{D_s}
\end{equation}
is the Einstein radius of the model. When the angular separation becomes similar to the Einstein radius, the magnification goes singular, producing an `Einstein ring'.

\begin{table*}[ht]
    \centering
    \caption{Estimated parameters of the best-fit mass density profile scenarios for each richness range.}
    \begin{tabular}{cccccccc}
    \hline\hline
    & & Bin 1 & Bin 2 & Bin 3 & Bin 4 & Bin 5 & Total\\
    \hline
    & $M_{SIS} [10^{13} M_\odot]$ & 0.5 & 0.6 & 0.6 & 0.6 & 1.0 & 0.5\\
    SIS+NFW & $M_{NFW} [10^{13} M_\odot]$ & 4.9 & 5.3 & 10.1 & 14.0 & 51.5 & 5.5 \\
    & $C$ & 0.94 & 0.30 & 1.17 & 0.65 & 0.56 & 1.84\\
    \hline
    %   & $M_{SIS} [10^{13} M_\odot]$ & 5.9 & 9.4 & 9.0 & 9.4 & 10.7 & 5.2\\
    Outer  & $M_{NFW} [10^{13} M_\odot]$ & 5.8 & 7.9 & 11.2 & 27.4 & 51.5 & 7.1 \\
    ($\gtrsim 100$ kpc) & $C$ & 0.74 & 0.39 & 1.00 & 1.74 & 0.56 & 1.72\\
    \hline
    %   & $M_{SIS} [10^{13} M_\odot]$ & 0.6 & 0.6 & 0.6 & 0.6 & 0.6 & 0.6 \\
    Inner & $M_{NFW} [10^{13} M_\odot]$ & 3.8 & 2.3 & 7.2 & 1.0 & 1.0 & 4.1\\
    ($\lesssim 100$ kpc)   & $C$ & 3.63 & 6.83 & 3.81 & 11.91 & 14.8 & 4.17\\
    \hline
    Inner + Outer & $M [10^{13} M_\odot]$ & 9.6 & 10.2 & 18.4 & 28.4 & 52.5 & 11.2\\
    \hline
    %\multicolumn{8}{c}{from catalogue}\\
    & $\langle R\rangle$ & 14.6 & 20.9 & 31.4 & 50.4 & 91.4 & 20.0\\
    From & $\langle M_{200}\rangle [10^{13} M_{\odot}]$ & 7 & 11 & 18 & 32 & 64 & 11 \\
    catalogue & $\langle z \rangle$ & 0.38 & 0.39 & 0.37 & 0.32 & 0.24 & 0.38\\
    & scale [kpc/"] & 5.42 & 5.51 & 5.33 & 4.85 & 3.96 & 5.42\\
    \hline\hline
    \end{tabular}
    \tablefoot{Estimated mass and concentration with the NFW+SIS and NFW profiles for the outer and inner parts and the total mass combining them (from top to bottom) for each richness bin and for the total case (from left to right). The last rows provide the average richness, mass, and redshift estimated from the catalogue and the corresponding scale factor.}
    \label{tab:datos}
\end{table*}

\section{Results}
\label{sec:results}

\begin{figure*}[ht]
\centering
 \includegraphics[width=0.4\textwidth]{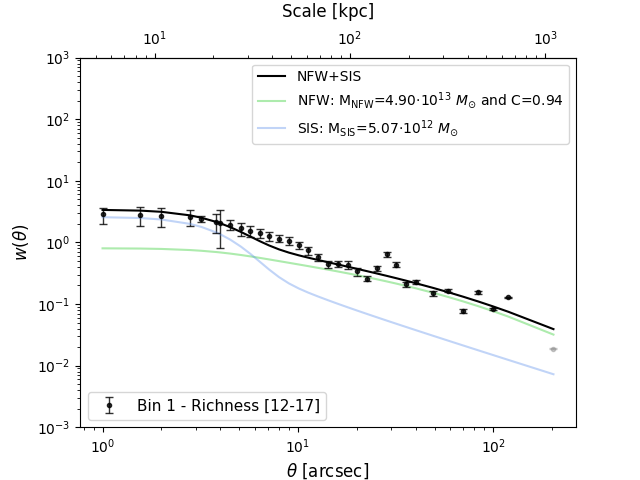}
 \includegraphics[width=0.4\textwidth]{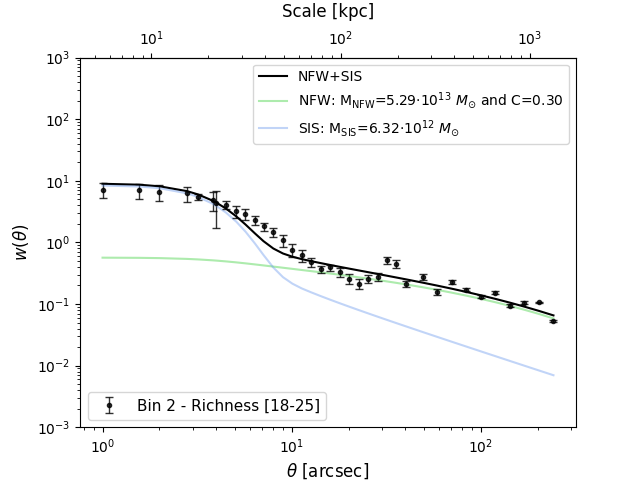}
 \includegraphics[width=0.4\textwidth]{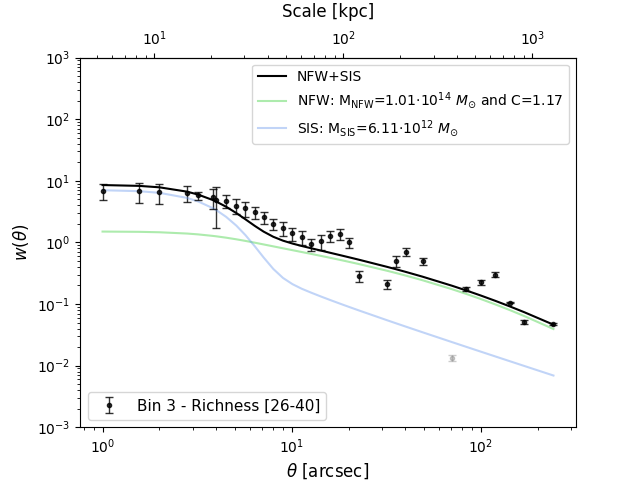}
 \includegraphics[width=0.4\textwidth]{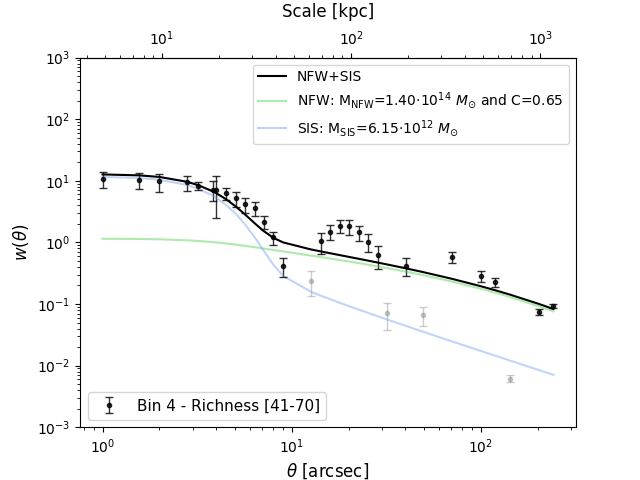}
 
\includegraphics[width=0.4\textwidth]{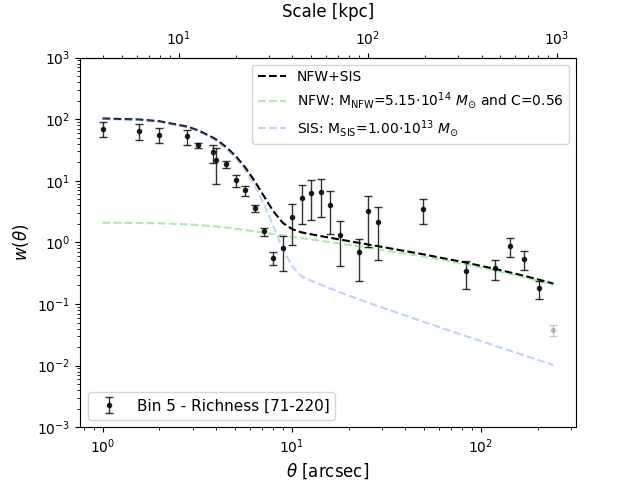}
 \includegraphics[width=0.4\textwidth]{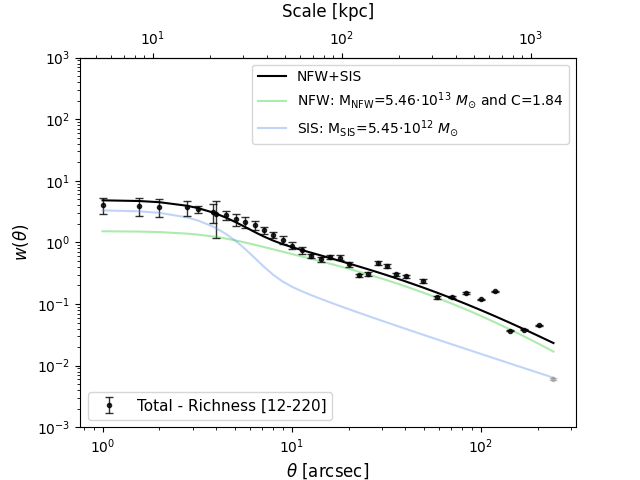}
 \caption{NFW+SIS fits (black line) to the cross-correlation data (black points) for bins 1 to 5 (from left to right, top to bottom) and the general case (bottom right). The corresponding SIS (blue lines) and NFW (green lines) fits are also separately shown. As discussed in more detail in sections \ref{sec:sisnfw} and \ref{sec:feature}, the best fits are always below the data for angular separations of between 5 and 10 arcsec and there is a potential lack of power at $\sim10$ and $\sim25$ arcsec. Grey points are considered outliers and are not taken into account for the analysis. The dashed lines for bin 5 indicates that the values were chosen by hand in order to produce a reasonable fit because the fitting algorithm does not converge (see text for more details).}
  \label{fig:SISSA_plots}
\end{figure*}

\subsection{Mass density profile fits}
\label{sec:profile_fits}
In order to analyse the measured CCFs and to extract physical conclusions, we try to fit the data to different combinations of the two common mass density profiles discussed above. In the theoretical modelling, the same smoothing as that applied to the data because of the positional uncertainty of the background sample was taken into account.

The different fits to the data produced in this work clearly show that a single mass density profile is unable to fit the data at all scales. Taking into account previous related analyses \citep[e.g. ][]{JOH07, BAU14, LAP12}, in this work we interpret the inner excess as the contribution of a galactic halo associated with the BCG, which is common in the centre of galaxy clusters. 

A summary of the best-fit values for the different masses and concentrations is shown in Table \ref{tab:datos}. Additional useful information from the cluster catalogue can also be found in the same table: mean redshift, richness, and mass for each richness bin and the total sample.

\subsubsection{SIS+NFW fit}\label{sec:sisnfw}
We first try to combine an SIS profile (a dark matter galactic halo plus a stellar contribution in the centre) to describe the BCG contribution plus a NFW for the contribution from the cluster halo (see Fig. \ref{fig:SISSA_plots}, black solid line). The black dots correspond to the CCFs obtained with stacking in each case.

We do not impose any angular scale restriction on either of the two profiles; it naturally arises from the different shapes of each profile. In addition, we notice that some bins show important fluctuations in the data with respect to the main trend. In order to derive reasonable fits, we decide to omit the clearest outliers (grey points in Figs. \ref{fig:SISSA_plots} and \ref{fig:NFW_plots}), that is, those measurements with a very low value compared to the adjacent ones. Otherwise, the fitting algorithm tries to take these points into account providing an unreasonable fit below the main trend indicated by the rest of the data. Removing additional data does not further affect the behaviour of the algorithm  and always provides the same results. See Sect. \ref{sec:feature} for a more detailed discussion on the potential physical interpretation of these fluctuations.

Moreover, the low number of CG pairs in the stacked images for bin 5 causes the strong oscillatory behaviour in the measured data. This issue prevents the fitting algorithm from converging and therefore the selected values are chosen by hand in order to produce a reasonable fit. This fact is indicated by the use of dashed lines in Figs. \ref{fig:SISSA_plots} and \ref{fig:NFW_plots}.

From the NFW profile (green lines), we find that the estimated masses for each richness bin increase monotonically  from M$_{\text{NFW}}=4.9\times 10^{13}M_{\odot}$ to M$_{\text{NFW}}=51.5\times 10^{13}M_{\odot}$ (M$_{\text{NFW}}=5.5\times 10^{13}M_{\odot}$ for the total sample). These values are always lower than the estimated ones for the mean richness in each bin, using the WHL12 mass-richness relationship (see Table \ref{tab:datos}). The retrieved concentration values do not show any clear trend with richness and they are much smaller than the expected values from the most common mass--concentration relationships \citep[e.g.][]{MAN08,DUT14,CHI18}. 

In the case of the SIS profile (blue lines), angular separations of a few arcseconds are close to the Einstein radius for the range of typical halo masses discussed in this work. As a consequence, both the magnification and the CCF diverge.
 
The subsequent required smoothing then produces a very particular mass density profile \citep[as already discussed in][]{BON19} that contributes only to the most inner data. 
Moreover, the derived masses, M$_{\text{NFW}}\sim 0.6\times 10^{13}M_{\odot}$ in all cases, are much smaller than the ones expected for a typical BCG. In fact, an effective halo mass of M$_{\text{eff}}=2.8-4.4\times10^{13}M_{\odot}$ is derived from the modelling of the large red galaxy (LRG) angular correlation function \citep{BLA08}. Similarly, consistent masses are also estimated from the analysis of their large-scale redshift-space distortions \citep[$M=3.5^{+1.8}_{-1.4}\times10^{13}M_{\odot}$, ][]{CAB09,BAU14}. BCGs are expected to have similar physical characteristics to the LRGs. 

Therefore, this combination of profiles, NFW+SIS, provides a good overall fit to the different set of data. For the total case, the fit correctly describes the data at all angular scales, but there is an issue at intermediate angular scales that is easy to identify when the sample is divided into different richness bins. For angular scales of between $\sim$5 and 10 arcsec, the fits are consistently below the data. 

\subsubsection{Inner and outer independent fits}
\begin{figure*}[ht]
\centering
 \includegraphics[width=0.4\textwidth]{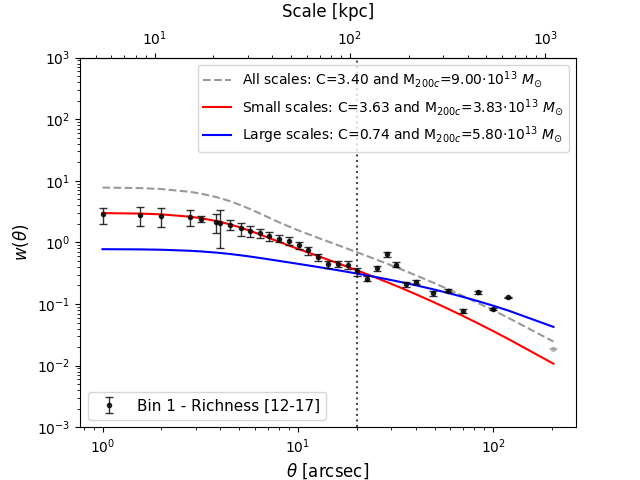}
 \includegraphics[width=0.4\textwidth]{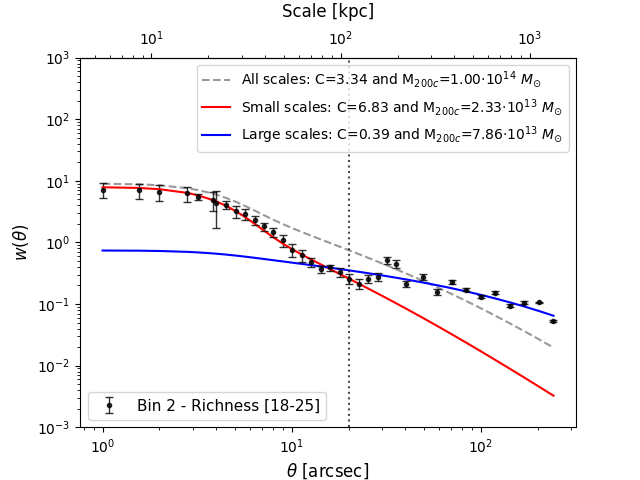}
 
 \includegraphics[width=0.4\textwidth]{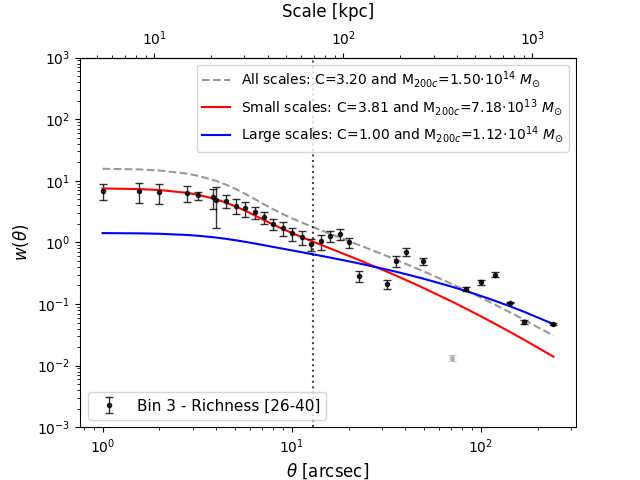}
 \includegraphics[width=0.4\textwidth]{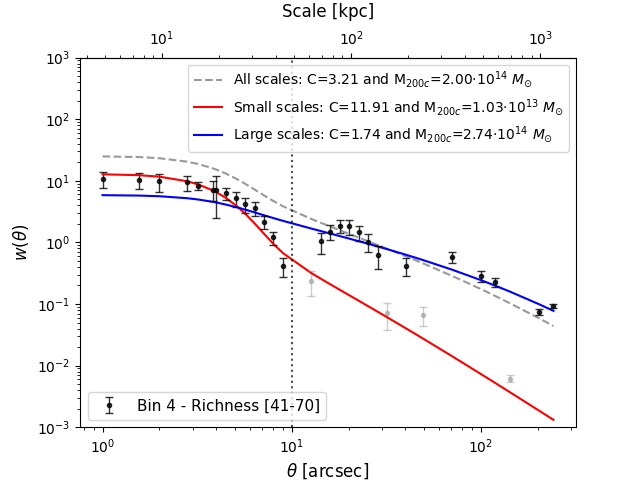}
 
\includegraphics[width=0.4\textwidth]{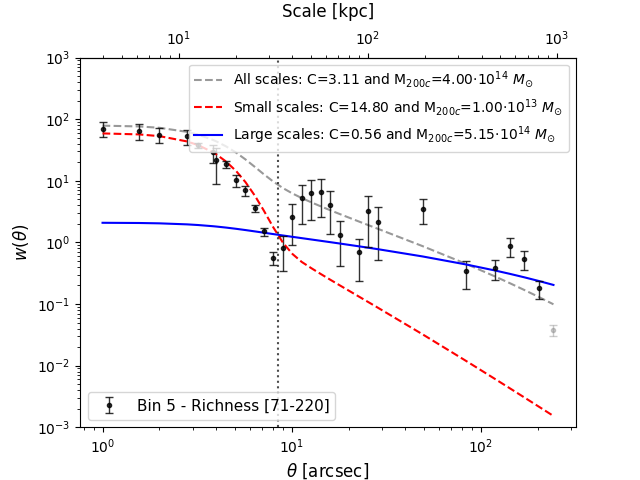}
 \includegraphics[width=0.4\textwidth]{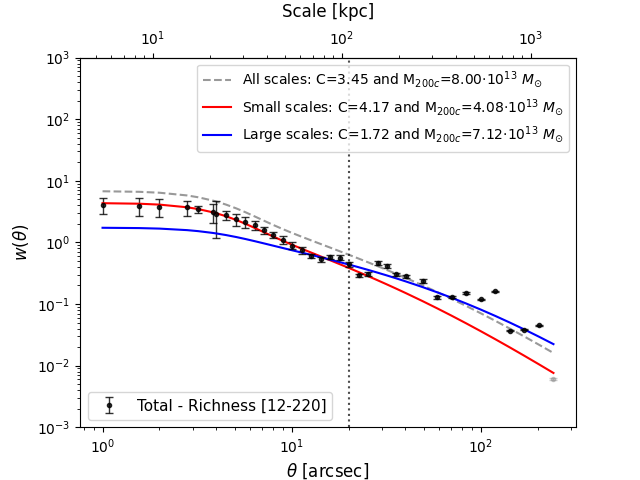}
 \caption{NFW fits to the cross-correlation data for bins 1 to 5 (from left to right, top to bottom) and the general case (bottom right). The red line corresponds to the fit to the points at the small scales only and the blue one is for large scales. The grey dashed lines show the NFW profile using the mass--concentration relationship by \citet{MAN08} and, as explained in the text, it is not a best fit to the data; neither is the red dashed line for bin 5 (the parameter values are chosen by hand in order to produce a reasonable fit). The vertical dotted lines separate the two regimes that have been studied for each case. Grey points are considered outliers and are not taken into account for the analysis.}
  \label{fig:NFW_plots}
\end{figure*}

For the reasons given above, we decide to perform independent analyses for the data at small (red line; inner part) and large (blue line; outer part) angular scales (see Fig. \ref{fig:NFW_plots}). In this case, we use a NFW for each of the regimes. We set the boundary (vertical dotted line in each panel) between the two regimes at around 10--20 arcsec ($\sim$ 52--105 kpc at z=0.38), where there is an unexpected increase in the CCFs. This feature is clearly seen for the highest richness bins but it is only a fluctuation in the total case.  

If we focus first on the large angular scales, we find that the estimated masses for each richness bin increase monotonically  from M$_{\text{NFW}}=5.8\times 10^{13}M_{\odot}$ to M$_{\text{NFW}}=51.5\times 10^{13}M_{\odot}$ (M$_{\text{NFW}}=7.1\times 10^{13}M_{\odot}$ for the total sample). Although compatible, these values are always lower than the estimated ones for the mean richness in each bin, using the WHL12 mass--richness relationship (see Table \ref{tab:datos}). The agreement improves towards higher richness. 

With respect to the concentration parameter, it also increases with richness, from $C=0.74$ to $C=2.22$ ($C=1.72$ for the total sample). However, these values are generally lower than the ones retrieved from the most common mass--concentration relationships \citep[e.g.][]{MAN08,DUT14,CHI18}. As a comparison, in Fig. \ref{fig:NFW_plots} we also plot (with a grey dashed line) the NFW profile using the one by \citet{MAN08}. It should be stressed that it is not the best fit to the large-scale data in each case, but serves to illustrate the different conclusions that would be drawn if it were used. Any fitting algorithm would have provided one of the following solutions: producing an excess at the smallest angular separations (as shown by the grey dashed lines) to fit the larger scales or completely underestimating the largest angular separations (similar to the inner fits shown as the red lines) to fit the smaller scales. In addition, the estimated masses derived using this mass-concentration relationship are higher than the ones obtained with the free fit. This issue becomes less relevant for higher masses or richness values.

The inner part is also well described by an NFW profile. In this case, the derived masses do not show any clear relationship with richness; the values fluctuate around M$_{\text{NFW}}=3-4\times10^{13}M_{\odot}$ (M$_{\text{NFW}}=4.1\times10^{13}M_{\odot}$ for the total sample). These halo mass values are in good agreement with previous independent LRG halo mass estimations, as discussed above. 

The derived inner concentration values are in better agreement with those expected from the mass--concentration relationships, with values around $C\sim 4$. 
This behaviour is opposite to that predicted by the mass--concentration relationships at higher halo masses. However, the lack of pairs in the highest richness bins makes this potential conclusion less reliable. Indeed, for bin 5, with just a single CG pair in the centre, the profile is completely dominated by the Gaussian kernel used to take positional uncertainties into account. The fit is completely degenerate (lower concentration values can be counterbalanced by higher masses) and we simply draw one of the possible fits with the lowest concentration value (again we use a dashed line to indicate that it is not a real best fit to the data). It is likely that, with better statistics, as in the other richness bins or the total case, the conclusion would be that the concentration remains almost constant, as expected for BCGs with similar masses.

Finally, it is interesting that by adding up the mass from the inner and outer fits (see `Inner+Outer' in Table \ref{tab:datos}) we get an almost perfect agreement with the average mass estimated from the WHL12 mass--richness relationship. We interpret this better agreement as validation of our procedure and as the necessity to use more complex mass density profiles in future works.

\subsection{Discussion}
\label{sec:discuss}
As presented in the previous section, our large-scale measurements ($\gtrsim 100$ kpc) are well described by an NFW profile, which is in agreement with a large number of previous studies. The derived masses are slightly lower but compatible to the WHL12 estimations, and therefore they also increase with richness, as expected.

In this respect, by comparing the CCF normalisation, we can approximately infer the average richness of previous works related to magnification bias. The CCF that is being used for cosmological analyses \citep[][]{GON17,BON20,GON21,CUE21} is measured using foreground samples built from galaxy catalogues. The values around 100 kpc are similar to the bin 1 or bin 2 measurements, which implies a richness of below 25. This is additional confirmation of the conclusion that is arrived at by these latter authors that the lenses that produce the magnification bias of the SMGs are not isolated massive galaxies but groups of galaxies or low-richness clusters. 

As in previous works \citep[e.g. ][]{BAU14,JOH07,OKA16}, we also find the necessity to include an additional central mass to explain our data. Although the use of a SIS profile helps to provide a good overall description of the data at all relevant angular scales, we find that the estimated masses are too low for BCGs and that the fit can be further improved at intermediate angular scales of $\sim5-10$ arcsec. By using a second independent NFW profile for the inner part, we find a better fit with an estimated central mass of M$_{\text{NFW}}=3-4\times10^{13}M_{\odot}$ that is more or less independent of richness. Therefore, these results confirm our assumption that this central mass corresponds to the presence of a BCG. 

Moreover, these halo masses are in agreement with the measured ones for massive LRGs. The fact that the inner data ($\lesssim 100$ kpc) are well described by an NFW profile implies that the lensing effect at these angular scales is dominated by the BCG galactic dark matter halo.  \citet{GAV07} arrived at similar conclusions based on a weak lensing analysis of 22 early-type (strong) lens galaxies, as did \citet{OKA16} based on an analysis of the central mass profiles of the nearby cool-core galaxy clusters Hydra A and A478. In both cases, the transition towards a central point like BCG stellar (baryonic) dominance is around 10 kpc, $<2$ arcsec for $z=0.38$ \citep[see also, ][]{LAP12}. These physical scales are beyond the current resolution of our measurements and therefore we can only observe the lensing effect of the galactic dark matter halo. This fact can also be related to the relatively poor fit using the SIS profile.

In addition, for the lowest richness bins ($R<40$), the central mass is $\sim 40$\% of the total mass. Therefore, these low-richness clusters are mainly composed of a massive central galaxy or BCG with several smaller satellite galaxies with 20 times lower masses  on average. For example, if we consider the total case, we have a central galaxy of M$_{\text{NFW}}=4.1\times
10^{13}M_{\odot}$ and the rest of the mass, M$_{\text{NFW}}=7.1\times10^{13}M_{\odot}$, is made up of the contribution from another 20 members, each of them with an average halo mass of $\sim 3.5\times10^{12}M_{\odot}$.

This result also confirms the conclusions from \citet{DUN20}. These latter authors observed an overdensity of high-redshift SMGs around a statistically complete sample of twelve 250 $\mu m$-selected galaxies at $z=0.35$, which were targeted by ALMA in a study of gas tracers. This observed overdensity is consistent with the magnification bias produced by halos of mass of the order of $7.1\times10^{13}M_{\odot}$, which are supposed to host one or possibly two bright galaxies and several smaller satellites. Indeed, of the six fields with unexpected SMGs, one is associated with a spectroscopically defined group and another four show observational evidence of an interaction between the central galaxy and the satellites.

Moreover, this scenario could also be related to the low concentration values obtained for our outer data. As already described in the previous section, all mass--concentration relations predict that the concentration should increase as halo mass decreases. However, these relationships are in general derived from the detailed analysis of individual clusters, which is only possible for the most massive ones \citep[e.g.][ with $M_{200c}\gtrsim 5\times10^{14}M_{\odot}$]{UME16}. As a comparison, our derived mass for the total case is at least seven to ten times smaller.

If we generalise the observational evidence obtained by \citet{DUN20}, we would expect  the group of galaxies, or equivalently the clusters of  lowest richness, to have lower concentration values. As halos are dynamically evolving objects, their mass and concentration is probably related to their recent assembly history \citep[][]{SER13}. Simulations show that in unrelaxed halos, much of the mass is far from the centre and therefore they tend to have lower concentrations than relaxed ones \citep[][]{CHI18}. At the same time, after a recent merger, the halo profile may not be well described by the NFW profile because of the dynamically unrelaxed state \citep[][]{CHI18}, although this problem should be mitigated in stacked halo profiles as in our case.
In addition, the fact that the cluster centre is determined as the position of the BCGs ---and not for example the centre of mass--- helps to obtain more homogeneous results even if a portion of the targets are in a dynamically unrelaxed state.
However, if this misalignment has become systematic and important for the chosen target sample, it can provide an additional smoothing factor that would limit the precision at the central region even for background samples with better positional accuracy.

\subsection{Lack of signal at $\sim 10$ and $\sim 25$ arcsec}
\label{sec:feature}

\begin{figure}[ht]
 \includegraphics[width=0.45\textwidth]{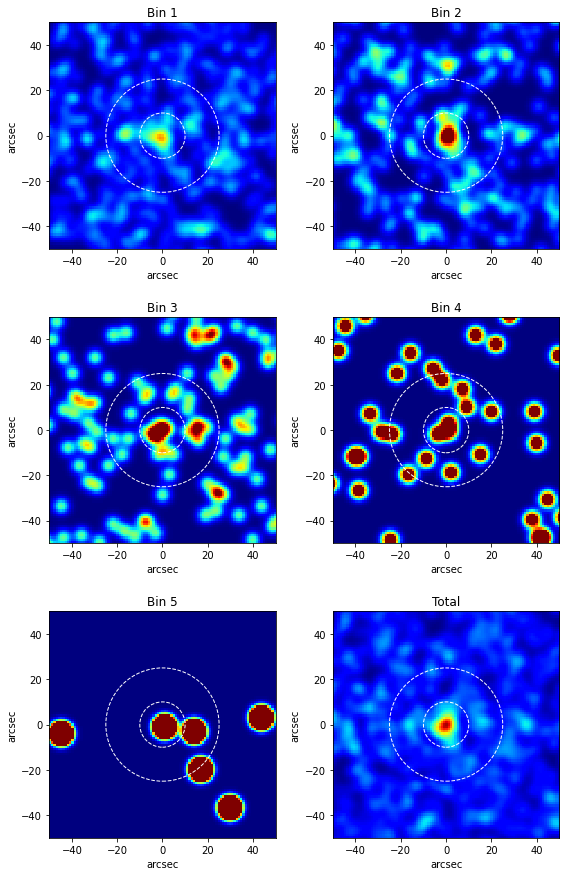}
 \caption{Highlight of the stacked maps of CG pairs for each richness bin at angular scales of lower that 50 arcsec. The maps are for bins 1 to 5 in the panels from left to right, and top to bottom. The bottom-right panel shows the highlight for the total case. The colour scale was chosen to improve visibility of the individual CG pairs and is the same for all panels. The two concentric circles (white dashed lines) indicate the radii of $\sim 10$ and $\sim 25$ arcsec.}
 \label{fig:bincircles}
\end{figure}

As already  mentioned, the large-scale fits have a number of issues and in some cases we do not consider certain data points in order to derive reasonable fits. At first sight, the issue seems to be related to the lack of CG pairs in the bins with the highest richness. However, we notice that this lack of signal at a certain angular separation is also present in all bins, and even in the total sample, but more subtly. From Fig. \ref{fig:NFW_plots}, we identify this issue at at  least two angular separations, $\sim 10$ and $\sim 25$ arcsec ($\sim 55$ and 125 kpc, respectively). This lack of signal in the profiles has to be produced by a lack of CG pairs in rings with such a radius and with a similar width to the angular resolution used in the radial profile. Figure \ref{fig:bincircles} is visual confirmation of the presence of such `rings'. 
The most central part of the stacked images for all the cases is shown in this figure, and we have plotted two concentric circles (white dashed lines) with radii of $\sim 10$ and $\sim 25$ arcsec.

We confirm that the presence of these features does not depend on the smoothing step, although their relevance and shape are affected for values of $\sigma$ greater than 5 arcsec, as expected. Considering that these rings can also be detected in the lowest richness bins, with hundreds of CG pairs, this indicates that they are not a statistical fluctuation, as could have been concluded simply from  bins 4 and 5.

Moreover, once this issue was recognised, we realised that a similar lack of signal was already present in previous works. In our recent studies, there is always an anomalous measurement (much lower than expected) around $\sim 30$ arcsec, which was chosen as the lowest angular separation for a weak lensing analysis using the CCF. This anomalous point is always present independently of the particular lens catalogue used: galaxies with spectroscopic redshifts from GAMAII \citep[][]{GON17,BON20,CUE21,DRI11}, SDSS galaxies with photometric redshifts \citep[][]{GON21}, or QSOs \citep[][]{BON19}. In addition, it can also
be found in the radial profile measurements using independent methodologies and/or catalogues: the weak lensing of the WHL12 cluster catalogue \citep[][]{BAU14}, the stacking analysis of the shear profile produced by galaxy clusters \citep[][]{JOH07}, and even the detailed joint analysis of strong lensing, weak lensing, shear, and magnification of individual galaxy clusters \citep[e.g. Abel209 or MACSJ0717.5+3745,][]{UME16}.

Interestingly, these angular scales correspond to the transition from  cluster dark matter halo dominance to BCG dark matter halo dominance. A similar behaviour was seen by \citet{GAV07} and \citet{OKA16} for the inner transition between the BCG galactic dark matter halo and the central stellar (baryonic) component: the measurements near the transition angular scales are below the theoretical expectation from the addition of both profiles.

Therefore, we conclude that the lack of signal in the transitions between different profile dominance regimes could be an indication of a physical phenomenon and not simply a statistical fluctuation. However, the detailed analysis and a potential physical interpretation of this effect are beyond the scope of this work.

\section{Conclusions}
\label{sec:concl}
In this study we exploited the magnification bias ---a gravitational lensing effect--- produced on SMGs observed by \textit{Herschel} at $1.2<z<4.0$ by galaxy clusters in SDSS-III with photometric redshifts of $0.05 < z < 0.8$ in order to analyse the average mass density profile properties of tens to hundreds of clusters of galaxies.

The measurements are obtained by stacking the CG pairs to estimate the CCF using the Davis-Peebles estimator. This methodology allows us to derive the mass density profile for a wide range of angular scales, $\sim$ 2-250 arcsec or $\sim$ 10-1300 kpc for $z=0.38$, with a high radial resolution, and in particular to study the inner part of the DM halo (< 100 kpc; the BCG does not impose any limitation as in other techniques). In addition, we also divide the cluster sample into 5 bins of richness.

Moreover, this methodology has some advantages from the point of view of analysis: it is straightforward to take positional uncertainties into account, which are critical at small angular separations, and to consider both the weak and strong lensing effects.

In order to completely describe the data for the full angular separation range, we need to take into account two dark matter halos (two different mass density profiles): a more massive halo to describe the outer part of the cluster ($>100$ kpc), and another halo for the inner part due to the presence of the BCG ($<100$ kpc). A good overall description is achieved by assuming a combination of a SIS profile (a dark matter galactic halo plus a stellar contribution in the centre) to describe the BCG contribution, plus an NFW profile to describe the contribution from the cluster halo. However, better results are derived for each regime individually using two independent NFW profiles. 

The average total masses (taking into account both NFW profiles) are in perfect agreement with the mass--richness relationship estimated by WHL12 (see Table \ref{tab:datos}). For the bins of  lowest richness, the central galactic halo constitutes $\sim 40$\% of the total mass of the cluster and its relevance diminishes as richness increases. While the estimated concentration values of the central galactic halos are in agreement with traditional mass--concentration relationships, we find lower concentrations for the outer part. Moreover, the concentrations decrease for lower richness, probably indicating that the group of galaxies cannot be considered relaxed systems.

Finally, we notice a systematic lack of signal at the transition between the dominance of the cluster halo and the central galactic halo ($\sim 100$ kpc). This is not deemed to be a statistical fluctuation or related to the smoothing step in the methodology pipeline. Moreover, this feature is also present in previous works using different catalogues and/or methodologies. Therefore, we conclude that it has a physical nature and merits a more detailed analysis. However, the physical interpretation of this lack of signal is beyond the scope of this paper and will be analysed in detail in a future study.

\begin{acknowledgements}
MMC, JGN, LB, DC, JMC acknowledge the PGC 2018 project PGC2018-101948-B-I00 (MICINN/FEDER).
MMC acknowledges PAPI-20-PF-23 (Universidad de Oviedo).\\
AL is supported by the EU H2020-MSCA-ITN-2019 Project 860744 “BiD4BESt: Big Data applications for black hole Evolution STudies.” and by
the PRIN MUR 2017 prot. 20173ML3WW “Opening the ALMA window on the cosmic evolution of gas, stars and supermassive black holes”.\\
We deeply acknowledge the CINECA award under the ISCRA initiative, for the availability of high performance computing resources and support. In particular the COSMOGAL projects “SIS20\_lapi”, “SIS21\_lapi” in the framework “Convenzione triennale SISSA-CINECA”.\\

The \textit{Herschel}-ATLAS is a project with \textit{Herschel}, which is an ESA space observatory with science instruments provided by European-led Principal Investigator consortia and with important participation from NASA. The H-ATLAS web-site is http://www.h-atlas.org. GAMA is a joint European-Australasian project based around a spectroscopic campaign using the Anglo-Australian Telescope.\\

This research has made use of the python packages \texttt{ipython} \citep{ipython}, \texttt{matplotlib} \citep{matplotlib} and \texttt{Scipy} \citep{scipy}.
\end{acknowledgements}

\bibliographystyle{aa} % style aa.bst
\bibliography{cluster_stack} % your references Yourfile.bib

\end{document}